\documentstyle[aps,preprint,floats,psfig]{revtex}
\begin{document}
\title{Specific heat in the thermodynamics of clusters}
\author{Peter Borrmann, Dorian Gloski, Eberhard R. Hilf}
\address{Department of Physics,
Carl v. Ossietzky University Oldenburg,\\
D-26111 Oldenburg, Germany}
%
%    A B S T R A C T
%
\maketitle
\begin{abstract}
The thermodynamic properties such as the specific heat are
uniquely determined by the second moments of the energy distribution
for a given ensemble averaging.
However for small particle numbers the results depend on the
ensemble chosen.
We calculated the higher moments
of the distributions of some observables
for both the canonical and the microcanonical ensemble of
the same van der Waals clusters.
The differences of the resulting thermodynamic observables
for the two ensembles are calculated in terms of the higher moments.
We demonstrate how for increasing particle number these terms
decrease to vanish for bulk material.\\
For the calculation of the specific heat within the microcanonical
ensemble we give a new method based on an analysis of histograms.
\end{abstract}
%
%      I N T R O D U C T I O N
%
\narrowtext
\section{Introduction}
Van der Waals clusters have been investigated
both with microcanonical and
canonical simulation methods in the past
(see \cite{berry,borr1,borr2,grimson}
and references therein). While canonical simulations are done almost
exclusively with the Metropolis algorithm
\cite{metropolis}, for microcanonical
simulations there are the alternatives of
doing them by molecular dynamics or
with the Creutz algorithm \cite{creutz1,creutz2}. \\
The main interest in  most publications so far lies on
the identification and classification
of phase transitions in van der Waals clusters.
The deficiencies of all simulation
methods mentioned above occur in the region of
the so-called {\sl Berry phase}. The
reason for that is the large number of isomers
which has to be taken into account
for a correct calculation of the partition function \cite{borr1}.\\
Albeit the fact, that we use for the reason of
good comparability Argon clusters as
our test system, in this work we will not be
concerned with the questions mentioned above.
Instead our main interest here lies in the
dependence of the higher moments of the
statistical distribution functions on the
particle number and  the differences
between microcanonical and canonical descriptions.
It is well known that for the case of the
canonical ensemble both the internal
energy $\langle E \rangle$ and the constant
volume heat capacity
$C_{\rm V} = \frac{1}{k_{\rm B} T^{2}}
\langle (E - \langle E \rangle)^{2} \rangle$
are approximately proportional to the number of degrees of freedom
(see e.g. \cite{reichl}). This immediately  sets up that the  relative
fluctuations in energy become  smaller as the system size increases.
\begin{equation}
\Delta_{\rm R}E \equiv
\sqrt{\langle (E - \langle E \rangle)^{2} \rangle} / \langle E \rangle
\sim N^{-1/2}
\end{equation}
Indeed one might see this as the
{\sl  reason} for the existence of the
Berry phase  and one  expects
that its temperature width should
decrease with increasing particle number.\\

The higher moments of  state variables are normally
not of much interest in
thermodynamics. The reason is that they can easily
be  calculated by means of the
dissipation-fluctuation theorem from the first moments.
For small systems, however, higher moments are a
unique tool to sensitively study
the subtle differences of the same thermodynamic
quantity  from partition
functions of different ensembles, since the higher
 moments explore more
sensitively the form of the distribution (see \cite{hill1,hill2}).

The van der Waals clusters explored here are an
 extremely simple example because
only one independent variable is to be given,
while the volume, the surface and so on adjust themselves.
%
%      C O M P U T A T I O N A L    M E T H O D
%
\section{Computational method}
A system of Ar$_{n}$ clusters is modelled with the usual
Lennard-Jones Potential for the interatomic binding,
\begin{equation}
v(r) = 4 \varepsilon \left( \left( \frac{\sigma}{r} \right)^{12}
     -                      \left( \frac{\sigma}{r} \right)^6 \right)
\end{equation}
with parameters $\varepsilon = 10.3 $meV
and $\sigma = 3.405 $\AA.
\subsection{Canonical Monte Carlo}
For the  canonical ensemble  the partition function
\begin{equation}
Z = \int \prod_{i=1}^{N} d\vec{x}_{i}
\exp( -\beta V(\vec{x}_{1}, \ldots, \vec{x}_{n}))
\end{equation}
is calculated using the well known Metropolis
\cite{metropolis} Monte Carlo
algorithm. For optimal performance  we use for the calculation of
all thermodynamic quantities the histogram method of Ferrenberg et al.
\cite {ferren1,ferren2,borr1}.
For convenience we give a short outline of the procedure.\\
 If we perform {\sl R} Metropolis simulations at parameters
$\beta_{i}$ and store the Monte Carlo data as histograms $N_{\beta}(E)$
with $n_{\beta_{i}}$ being the total number of observations in the
i'th run, the probability distribution is given through
\begin{equation}
P_{\beta}(E) = N_{\beta}(E) /n_{\beta} = W(E) \exp( -\beta E + \beta F),
\end{equation}
where $W(E)$ is the density of states, and $F$ is the Helmholtz free
energy. Following Ferrenberg et al. the normalized probability
distribution can be improved by
\begin{equation}
D_{\beta}(E) = \frac{\sum_{i=1}^{R} N_{i}(E)}{\sum_{i=1}^{R} n_{i}
\exp(-(\beta_{i} - \beta) E + \beta F_{i})}
\label{nor1}
\end{equation}
where
\begin{equation}
\sum_{E} D_{\beta_{i}} = \exp(\beta F_{i}),
\label{nor2}
\end{equation}
which is simply the partition function.
The values of $F_{i}$ can be determined within an additive by
selfconsistent iteration of (\ref{nor1}) and (\ref{nor2}).
Simply spoken, this method improves D$_{\beta}(E)$ by use
of simulation data at other histogram data weighted by the number
of observations. We slightly improved the method by using fitted
probability distributions which are sacrificed by $\chi^{2}$ tests.\\
Now the mean value of any  observable $O(E)$ can be calculated easily
as a function of $\beta$.
\begin{equation}
\langle O_{\beta}(E) \rangle = \frac{1}{Z_{\beta}}
\int_{E} O(E) D_{\beta}(E)  \; .
\end{equation}
\subsection{Microcanonical Monte Carlo}
The microcanonical partition function
\begin{equation}
Z = \int \prod_{i=1}^{n} d\vec{x}_{i} \prod_{j=1}^{n} d\vec{p}_{j} \;
\delta[ H( \vec{x}_{1},\ldots,\vec{x}_{n},
\vec{p}_{1}, \ldots, \vec{p}_{n}) - E]
\end{equation}
is approximated with the microcanonical Monte Carlo algorithm invented
by Creutz \cite{creutz1,creutz2,challa}. The algorithm simulates the
integral
\begin{equation}  \label{demonpart}
Z = \int \prod_{i=1}^{n} {\rm d}\vec{x}_{i}
\int_{0}^{E-V_{0}} dE_{\rm D}
\; \delta[ V( \vec{x}_{1},\ldots,\vec{x}_{n}) + E_{\rm D} - E].
\end{equation}
where V$_{0}$ is the minimal potential energy,
in our case the potential energy of
the {\sl best single cluster} configuration.
E$_{\rm D}$ is an extra degree of
freedom called demon, which simulates the kinetic
energy of the system and is
restricted to positive values. As in the Metropolis algorithm
new configurations ${\bf x'}$ are chosen  at random.
Here the new configuration
is accepted if
\begin{equation}
\Delta V = V({\bf x'}) - V({\bf x}) <  E_{\rm D}.
\end{equation}
In this case ${\bf x'}$ is counted as a new configuration and the demon
is set to $E_{\rm D} \leftarrow E_{\rm D} + \Delta V$, otherwise
the step is rejected.
Pictorially the demon might be viewed as a tiny thermometer
thrown into  a large
swimming pool. If we denote the cluster system by A$_{\rm C}$,
the demon system
by A$_{\rm D}$ and the combined system by A$^{0}$ , the conservation
of energy can
be written as $E_{\rm C} + E_{\rm D} = E^{0}$.
The expansion of $\ln Z$ in
a taylor series yields
\begin{eqnarray}
\ln Z_{\rm C}( E^{0} -E_{\rm D}) &=&
\ln Z_{\rm C}(E^{0}) + \\ \nonumber
 &&\sum_{i}^{\infty}
\frac{(-1)^{i}}{ i !} \left[ \frac{\partial^{i} \ln Z_{\rm C}}
{ \partial E_{\rm C}^{i}} \right]_{E^{0}} E_{\rm D}^{i} .
\end{eqnarray}
While the first derivative is easily identified as $\beta$
\begin{equation}
\left[  \frac{\partial \ln Z_{\rm C}}
{ \partial E_{\rm C}} \right]_{E^{0}} \equiv \beta
\end{equation}
the higher derivatives are in turn derivatives of $\beta$.
The probability distribution for the demon energy is now given
by
\begin{equation} \label{probab}
P(E_{\rm D}) = C \exp\left( - \beta E_{\rm D} + \frac{1}{2}
\left[ \frac{\partial \beta}{\partial E} \right]_{E_{0}} E_{\rm D}^{2}
\ldots \right)
\end{equation}
In the case that the demon energy is sufficiently small compared to the
total energy of the system ,  $\beta$  is  indeed
the inverse of the demon energy as stated in the literature
\cite{creutz1,creutz2}.
\begin{equation}
\langle E_{D} \rangle =  \int_{0}^{E^{0}-V_{0}} dE_{D} \;  E_{D}
P(E_{D}) = \frac{1}{\beta}
\end{equation}
%
%     M U L T I P L E    N O R M A L     M O D E S
%
\subsection{Multiple normal modes}
In the multiple normal modes (MNM) model, described in detail in
\cite{borr2}, we take into account several isomers of a cluster,
characterizing each isomer by it's binding energy, permutational
degeneracy, and normal modes spectrum. \\
The ensemble partition functions are constructed from the single
isomer partition functions with proper ensemble dependent weights.\\

For the statistical equilibrium of isomers, the calculated
one-isomer partition functions have to be multiplied by
a factor $\sigma_{i}$ reflecting the permutational degeneracy
R$_{i}$.
In order to relate all of them to a common energy
(all particles free) also the exponential of the
binding energy $E_i$ appears as a relative weight between
configurations in the canonical partition function
\begin{equation}
Z = \sum_i \sigma_i e^{-\beta E_i} Z_i  \hspace{3mm} .
\end{equation}

Within the normal modes analysis for a given isomer the potential
energy is expanded up to second order around the ideal equilibrium
position of the isomer.
\begin{equation}
V({\bf x}) = \sum_{\alpha > \beta} v(r_{\alpha \beta})
             \approx \frac{1}{2} \sum_{\alpha \beta i j}
             P_{\alpha \beta}^{i j} \, x_{\alpha}^i \, x_{\beta}^j
\end{equation}
where $x_{\alpha}^{i}$ be the $i$'th spatial
component of the position of the particle $\alpha$ with respect to
its ideal equilibrium position, $x_{\alpha}^i = 0$.
Diagonalization of the  matrix {\bf P}  yields the 6N-6
eigenfrequencies \footnote{Six eigenvalues vanish due to zero
total momentum and total angular momentum. (In the case of fully
linear structures one has 6N-5 eigenfrequencies.) }.\\
All investigated isomers were copied from
``snapshots'' of MC-calculations or constructed ``by hand'' and
relaxed numerically until the internal forces were smaller than
$10^{-12} \varepsilon / \sigma$. Then the matrix $\bf P$
was calculated and diagonalized numerically.
With $\omega_k^i$ being the $k$-th eigenfreqency of the $i$-th
isomer, the canonical partition function is
\begin{equation}  \label{nmncanon}
Z = \sum_i \sigma_i e^{\beta E_i}
    \prod_{k=1}^{3N-6} \frac{2 \pi}{\beta \omega_k^i} \quad .
\end{equation}

For the microcanonical partition function
at first one has to calculate the micro-canonical
MNM-partition function for one isomer from the phase space integral:
\begin{eqnarray}
Z^{\mu}_i(E) &=&   \int ({\rm d}p \; {\rm d}q)^{3N-6} \;
                   \delta(\varepsilon (q,p) - E)\\
             &=&   O_{6N-12} \frac{1}{2E}
                   \prod_{k=1}^{3N-6} \frac{2E}{\omega_k^i}
\end{eqnarray}
where $O_n$ is the surface of the $n$-dimensional unit-sphere.
To obtain the MNM-partition function, all isomer-partition functions
have to be related to a common reference-energy, and summed up
% relative for respective
with the relative permutational degeneracies $\sigma_i$ ,
\begin{equation} \label{mnmmicro}
Z^{\mu}(E)   =   \sum_i \sigma_i Z^{\mu}_i(E - E_i).
\end{equation}
 From (\ref{nmncanon}) and (\ref{mnmmicro}) now all thermodynamic
properties can be calculated.
%
%       R E S U L T S
%
\section{Results}
\begin{table}
\caption{Total binding energy  and  relative degeneracies
of the most important Ar$_{13}$ isomers.}
\begin{tabular}{crr}
Isomer & $E_{\rm bind} / \varepsilon$ & $R$  \\
\hline
pure icosahedron        & 44.33 &    1  \\
singly decorated        & 41.47 &  180  \\
doubly dec., neighb.    & 40.62 &  900  \\
doubly dec., dist.      & 39.71 & 4800  \\
\end{tabular}
\end{table}
\begin{figure}
\centerline{\psfig{figure=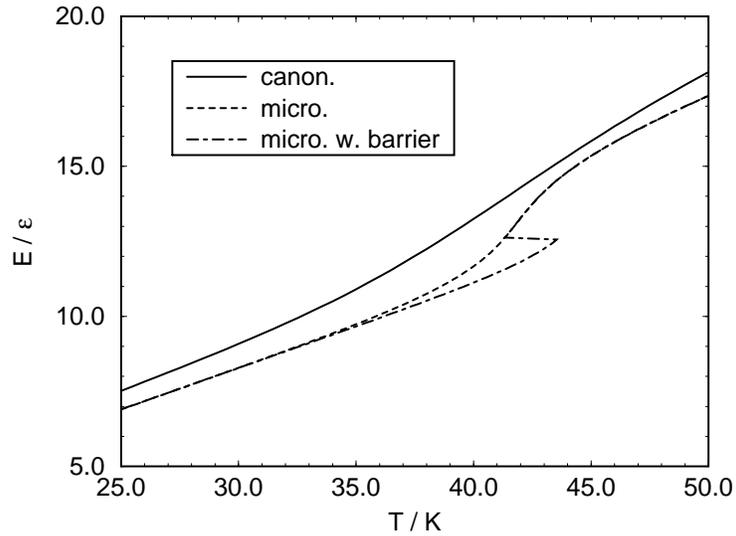,height=3.0in,angle=270}}
\caption{Canonical versus microcanonical caloric curve of Ar$_{13}$
calculated within the MNM model. The inclusion
of a virtual activation barrier between the icosahedral and the singly
decorated isomer of about 60 meV  in the microcanonical model
results in a van der Waals loop which is often encountered in
molecular dynamics simulations.}
\end{figure}
\begin{figure}
\centerline{\psfig{figure=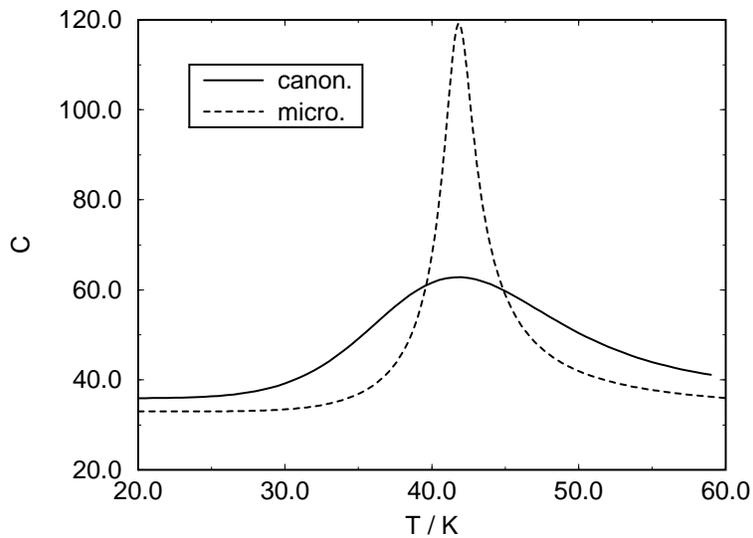,height=3.0in,angle=270}}
\caption{Canonical versus microcanonical specific heat of Ar$_{13}$
calculated within the multiple normal modes model.}
\end{figure}
\begin{figure}
\centerline{\psfig{figure=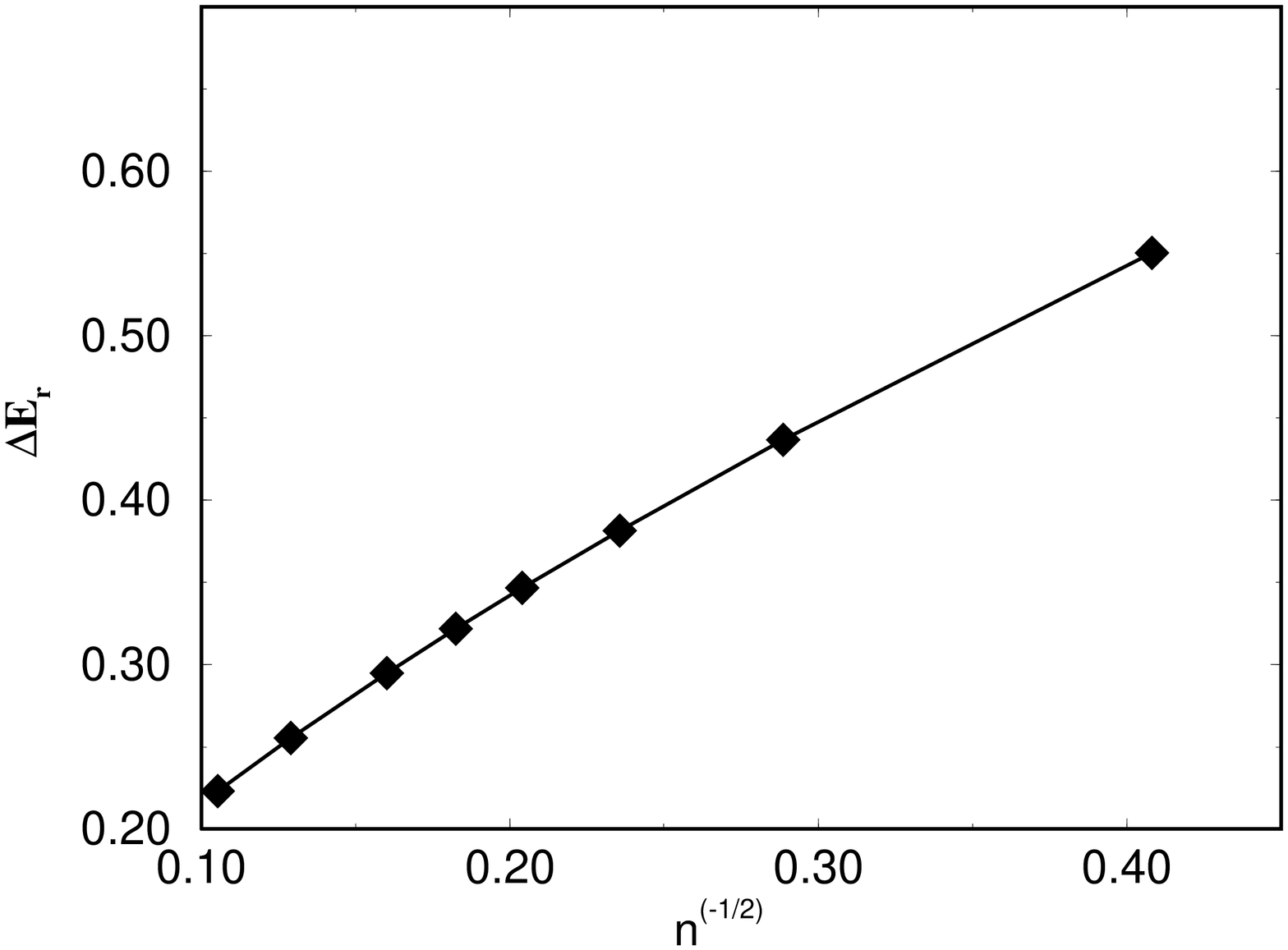,height=3.0in,angle=270}}
\caption{Dependence of the relative energy fluctuation on the
number of degrees of freedom calculated with canonical
Monte Carlo at T $\approx$ 5 K.}
\end{figure}
The multiple normal modes model gives a unique
chance to study the differences
between microcanonical and canonical ensembles, because for both ensembles
all thermodynamic quantities can be calculated exactly.
Fig.1 shows the caloric curves for Ar$_{13}$ clusters.
In our model calculations the
four most important isomers of Ar$_{13}$,
for  which  the binding  energies and relative
degeneracies are given in Table 1, have
been included. Both curves have a similar form
but differ significantly by a certain amount of energy.
A quite nice feature of the model is that the so-called
van der Waals loops, often encountered in molecular
dynamics simulations, can
qualitatively be reproduced by insertion
of an activation barrier into the
microcanonical model, i.e. by reducing
the accessible phase space up to a
specified energy barrier.
The differences between the ensembles get more drastical
if one examines the specific heat ( see Fig. 2).
The phase transition occurs in both
descriptions at the  same temperature
but is much sharper in the microcanonical ensemble.\\
Besides the deficiency, that within the
multiple normal modes model only
harmonic excitations of each isomer are
considered, it is very difficult to find all
important isomers and  their permutational degeneracy
as the cluster size increases.
In Monte Carlo calculations these features are
included automatically, but they
are exact only within the limit of infinite computation time.\\
Fig.3 shows the dependence of  the relative energy fluctuation (1) as a
function of the cluster size at a
temperature of $\approx$ 5 K calculated
from canonical Monte Carlo simulations.
As expected, / $\Delta_{\rm R}$E is
approximately proportional to the inverse of
the square root of the particle number. \\
\begin{figure}
\centerline{\psfig{figure=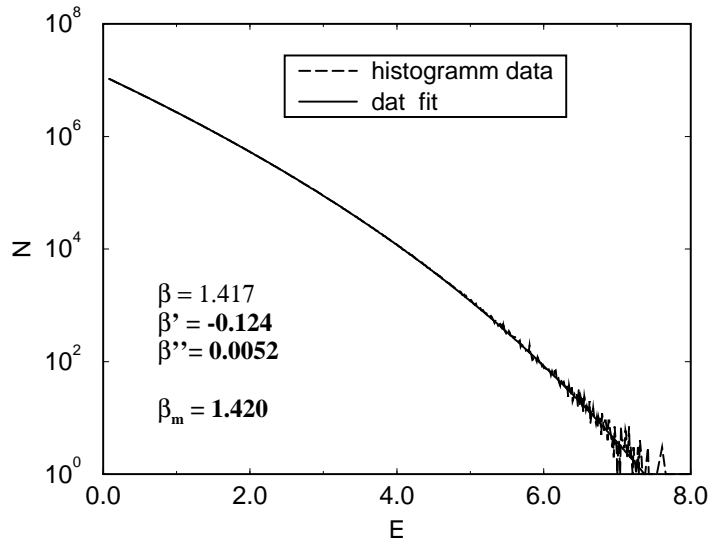,height=3.0in,angle=270}}
\caption{Fit of the demon energy histogram
data to the theoretical probability
distribution function (13). The histogram was obtained for an Ar$_{13}$
cluster at  E = -445 meV. $\beta_{m}$ is
obtained from the mean value equation (13).}
\end{figure}
To get comparable results between microcanonical
and canonical simulations a
primary task is to calculate the temperature within  the microcanonical
simulations. The simple choice to do
that is, of course, making use of equation (14)
by neglecting the higher order terms. Besides that we tried
another way, which enables us to calculate the
derivatives of $\beta$, too  by
fitting the histogram data of the
demon energy to the probability function (\ref{probab}).
Fig. 4 displays  an example of such a fit.
Albeit the difference between the
calculated $\beta$'s is quite small it is not negligible.
By chance we have found
a cheap way to calculate the derivatives of $\beta$.\\
\begin{figure}
\centerline{\psfig{figure=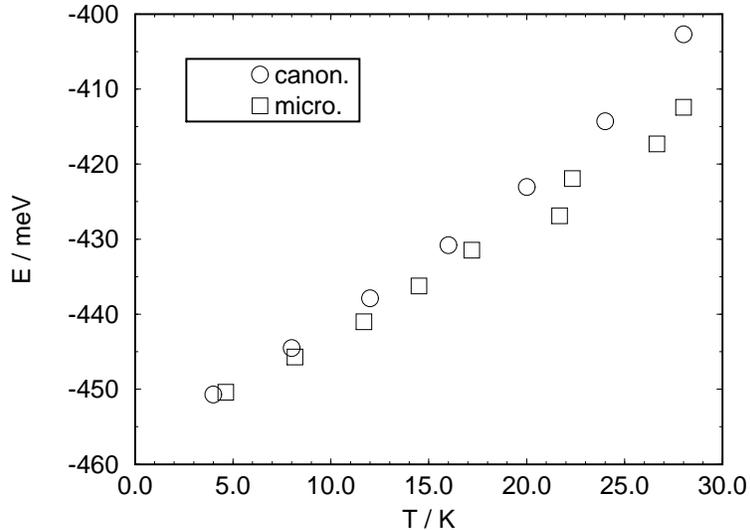,height=3.0in,angle=270}}
\caption{Caloric curve for Ar$_{13}$ from microcanonical and canonical
Monte Carlo simulations. Every data point was evaluated with $8*10^{7}$
steps.}
\end{figure}
The caloric curves of our Monte Carlo
simulations reveal a behavior very
similar to that found within the MNM model (see Fig. 5).
Preliminary  results show  that the relative fluctuations
of the temperature
in the microcanonical ensemble decreases similar
to the relative energy fluctuation shown in
Fig. 3 for the canonical ensemble.
It is very difficult to compare the
results for the higher moments of the
different ensemble quantitatively because the differences
depend not only on
the particle number but also very strong
on the temperature  (see Fig.2.).
Probably other system with a smaller transition
phase, e.g. Ising systems,
are more appropriate for a quantitative study of such dependencies.
Nonetheless the decrease of the relative
fluctuations of the state variables
with increasing system size in both
ensemble indicates that the differences
between the ensembles decrease with $\approx N^{-1/2}$ too.

%
%      C O N C L U S I O N S
%
\section{Conclusions \& Outlook}
We have shown that in the case of small systems it is not  unimportant
which  thermodynamical ensemble is used to describe the clusters. \\
 Things might get more interesting if
more state variables are considered, e.g.
by introducing spin degrees of freedom
along with a coupling to an external
magnetic field. A study for such a system  is in preparation.
To investigate the size dependencies
more properly not only the range of
the cluster size (in this study  up to 40) has to be expanded.
Because of possible magic number effects in every size region some
neighbor numbers should be
considered in order to average out such effects.\\
The accurate calculation of higher moments, or derivatives of $\beta$
within the microcanonical ensemble,
remains  a problem to be solved. One
solution could be to invent a procedure similar to the optimized data
analysis of Ferrenberg.
%
%      R E S U L T S
%

%
%    F i G U R E S
%
%
%     T A B L E S
%
\end{document}